\documentclass{aa}
\usepackage{graphicx,txfonts}
\usepackage{natbib}
\bibpunct{(}{)}{;}{a}{}{,}

\begin{document}

\title{On  the Lyman~$\alpha$ and Lyman~$\beta$ lines in  solar coronal streamers}

\author{N. Labrosse \and X. Li \and B. Li}

\offprints{N. Labrosse}

\institute{Institute of Mathematical and Physical Sciences, University of Wales Aberystwyth, Ceredigion, SY23 3BZ, UK\\
\email{nll@aber.ac.uk,xxl@aber.ac.uk,bbl@aber.ac.uk}}

\date{Received 21 February 2006 / Accepted 10 May 2006}

\abstract
{}
{We investigate the formation of the \ion{H}{i} Lyman~$\alpha$ and Lyman~$\beta$ lines in an equatorial coronal streamer. Particular attention is paid to frequency redistribution for the scattering of the incident radiation. The properties of the spectral lines are studied.}
{The coronal model is given by a global 2.5-D three fluid solar wind model with $\alpha$ particles. The emergent intensities and line profiles are calculated from the solution of the statistical equilibrium and radiative transfer equations for an hydrogen atom with 11 energy levels under non local thermodynamic equilibrium. The formation of the lines results from radiative excitation, collisional excitation, and takes into account the coupling with all other transitions between the hydrogen energy levels.}
{We present new estimates of the radiative and collisional contributions of the Lyman line intensities within the streamer. It is also shown that within the streamer, the full width at half-maximum (FWHM) of the Lyman~$\beta$ line is a better indicator of the plasma temperature than that of Lyman~$\alpha$.
These results show that care should be taken when inferring the proton temperature from the Lyman~$\alpha$ line profile as observed in coronal streamers, e.g. by the \emph{Ultraviolet Coronagraph Spectrometer} or the \emph{Solar Ultraviolet Measurements of Emitted Radiation} experiments on board the \emph{Solar and Heliospheric Observatory}.}
{}


\maketitle

\section{Introduction}

A large amount of our knowledge of the solar extreme ultraviolet (EUV) corona comes from the observations of the first two \ion{H}{i} Lyman lines at 1215.67 \AA\ and 1025.72 \AA. The Lyman~$\alpha$ line is indeed the brightest line in this range of radiation. To infer the coronal plasma properties from observations it is necessary to get the line profiles with enough spectral resolution. 
{While chromospheric profiles show a self-reversal at line centre, the coronal profile of the scattered Lyman~$\alpha$ line is close to a gaussian profile.}

From the knowledge of the line width one can derive an {equivalent} temperature {(ie. including all causes of line broadening),} and with further assumptions, the kinetic {(thermal)} temperature of the hydrogen atoms. If the coupling due to charge exchange between hydrogen atoms and ions is strong enough, then the proton temperature is equal to the hydrogen temperature. 
{\cite{loaetal98,loaetal00} concluded from fast solar wind models that this coupling is strong up to 3 R$_{\sun}$, while \cite{olh94} studied slow wind models and found strong coupling up to 10 R$_{\sun}$ in some conditions. The decoupling of hydrogen atoms and protons will occur at different densities, depending on which physical assumptions are made in the models. In our case, this coupling is strong enough at densities above 10$^6$ cm$^{-3}$. This condition is met within the whole streamer \citep[see for further details][]{lll06}.}
Thus we can obtain a reliable estimate of the proton temperature from the full width at half maximum (FWHM) of the Lyman line profiles \citep[see also][]{marschetal99}.  

In this paper we show that in the case of coronal streamers, radiative transfer calculations in non local thermodynamic equilibrium (NLTE) are useful to predict the properties of the Lyman~$\alpha$ and Lyman~$\beta$ lines. The temperature diagnostic from these two lines is examined. We describe the streamer model in section~\ref{bomodel} and the radiative transfer calculations in section~\ref{radtrans}. Results are discussed in section~\ref{discuss}.

\section{The coronal model}\label{bomodel}

\begin{figure}
  \centering
  \resizebox{\hsize}{!}{\includegraphics{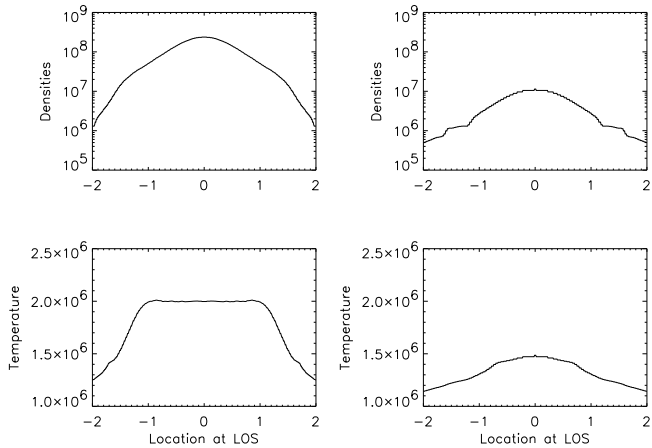}}
  \caption{Streamer models at a distance of 0.05~R$_{\sun}$ (left) and 0.88 R$_{\sun}$ (right) above the solar surface. Top panels: proton density in cm$^{-3}$; bottom panels: electron temperature in K. All quantities are plotted against the location along the line-of-sight expressed in units of the solar radius R$_{\sun}$, where the origin is at the centre of the LOS.}
  \label{streamer}
\end{figure}

The coronal model is obtained from a global three fluid solar wind model with $\alpha$ particles. It is described in another paper by {\cite{lll06}}. {It is a 2.5-D, axially symmetric model, where all variables depend on two spatial coordinates, but where the three components of vector quantities are retained. }In the streamer (the closed magnetic field region) no external heating is applied. A hot coronal boundary, electron heat flux and Coulomb coupling lead to a non-isothermal streamer in which all three species (namely electrons, protons, and $\alpha$'s) have the same temperature. The properties of the streamer model that we use for this study are shown in Fig.~\ref{streamer} for two heights. One has the line-of-sight (LOS) centre situated at a heliocentric distance of 1.05 R$_{\sun}$ at the equator (left column) and the other one is at a heliocentric distance of 1.88 R$_{\sun}$, slightly below the cusp which is located at $\sim 2$~R$_{\sun}$. The LOS is chosen to be 4~R$_{\sun}$ long. The top panels in Fig.~\ref{streamer} present the density variations along the LOS of the protons. The electron density is almost the same as the proton density due to the small amount of $\alpha$'s. The bottom panels of the figure show the temperature variation along the LOS.
For the present investigation we have made use of 10 different LOS for a streamer axis along the equator. All LOS are 4~R$_{\sun}$ long, but the centre of the LOS is located at an increasing distance from the sun. As one goes further away from the surface, the temperature profiles flatten  and the density variations become smoother as well as less sharp.

\section{Computation of the emergent radiation}\label{radtrans}

In order to compute the radiation emitted by the corona we use a numerical code which solves the radiative transfer (RT) equations and the statistical equilibrium (SE) equations in NLTE. This code has been described in \cite{ghv,gl00,gh02}. Here we recall its most relevant features.
The SE is solved for a 10 level + continuum hydrogen atom. The RT is solved for the lines and continua in a 1D plane-parallel geometry using a Feautrier method \citep{feautrier} with variable Eddington factors. The boundary conditions for the RT equations are determined by the radiation coming from the disk. This incident radiation is allowed to freely penetrate the structure. The Lyman~$\alpha$ profile from the solar disk is taken from OSO-8 observations \citep{glva78}. For Lyman~$\beta$ to Lyman-9 they are taken from SUMER observations \citep{warrenetal98}. Thomson scattering and Rayleigh scattering are taken into account in the computation of the continuous absorption coefficients.

\begin{figure}
  \centering
  \resizebox{\hsize}{!}{\includegraphics{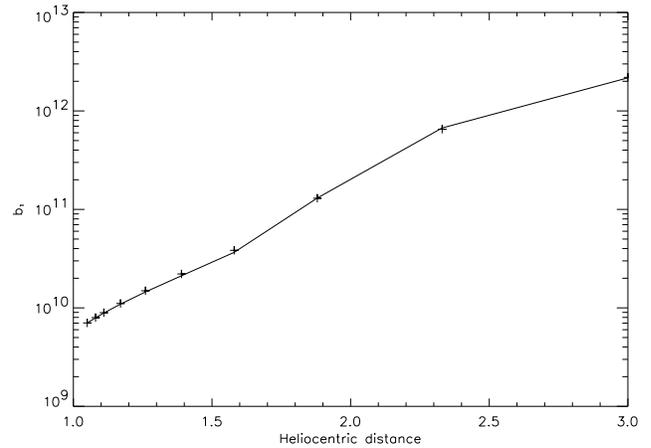}}
  \caption{LTE departure coefficient for the ground level of hydrogen as a function of heliocentric distance (solid line). The 'plus' signs give the value of $1/N_e$ in arbitrary units (see text).}
  \label{bih}
\end{figure}

It is necessary to perform the calculations in NLTE. Because of the importance of the incident radiation emitted by the solar disk and the low coronal densities, LTE cannot be reached. We define the LTE departure coefficient $b_i$ of the energy level $i$ by:
\begin{equation}
  b_i = \frac{N_i}{N_c} \left[ \frac{N_c}{N_i} \right] _\mathrm{LTE} . \label{ltedc}
\end{equation}
With this definition, the LTE departure coefficient is $b_c=1$ for the continuum level $c$. Here $N_i$ and $N_c$ are the populations of the bound level $i$ and continuum $c$. $\left[ \frac{N_c}{N_i} \right] _\mathrm{LTE}$ is given by the Saha-Boltzmann distribution.
Figure~\ref{bih} presents the value of the LTE departure coefficients $b_1$ of the ground level of hydrogen for our 10 LOS. It shows that the ground level is far from LTE, and that this tendency increases with height. This can be explained by the fall-off in densities, which makes the incident radiation even more predominant in the formation of the hydrogen spectrum at higher altitudes. In fact the variation of $b_1$ closely follows that of $1/N_e$ (inverse of the mean electron density) with altitude, where the mean of the density is defined as $ N_e = \int_0^L{n_e(x) \mathrm{d}x}/L$,
with $L$ the total length of the LOS. This is illustrated in Fig.~\ref{bih} where we plotted as 'plus' signs $1/N_e$ in arbitrary units normalized so that the values of $b_1$ and $1/N_e$ at 1.05 R$_{\sun}$ are identical.
{The factor of proportionality between $b_1$ and $1/N_e$ is dependent on temperature. This result is valid as long as the ionization balance of hydrogen is governed by collisional ionization and radiative recombination.}
The excited levels also are far from LTE but to a lesser extent. For instance we have $b_2/b_1 \lesssim 10^{-8}$.

Once the coupled system of SE and RT equations is solved we compute the emergent intensity in a particular line from the knowledge of the line source function, the source function for continuous absorption, and the related absorption coefficients. {The emergent intensity $I_\nu(\mu)$ at an angle $\theta$ between the normal to the surface and the LOS such that $\mu=\cos\theta$ is given by:
\begin{equation}
  \label{solform}
  I_\nu(\mu) = \int_{0}^{\tau_\nu}{S_\nu(t) \mathrm{e}^{-t/\mu} \ \mathrm{d}t/\mu} \ ,
\end{equation}
where $\tau_\nu$ is the optical {thickness} at frequency $\nu$.}
Of course the simplicity of Eq.~(\ref{solform}) hides the fact that {the computation of the total source function $S_\nu(\tau_\nu)$ in NLTE is a non-trivial task.} Here we use the formulation of the equivalent two-level atom. One can then express the line source function as:
\begin{equation}
  \label{source}
  S_\nu^l = \varepsilon^\ast B^\ast + (1-\varepsilon^\ast) \tilde{J}_\nu .
\end{equation}
In Eq.~(\ref{source}), $\varepsilon^\ast$ and $B^\ast$ account for all the processes that can affect the creation and the destruction of photons in the transition at frequency $\nu$, while $\mathbf{(1-\varepsilon^\ast)} \tilde{J_\nu}$ is the {\emph{scattering} term of the source function.} Expressions for $\varepsilon^\ast$ and $B^\ast$ are given in \cite{phdgoutte} \citep[see also][Chap.~8.1]{jeff}. Through Eq.~(\ref{source}) a non-local and non-linear coupling between the radiation and the plasma arises.

{The scattering {integral} in eq.~(\ref{source}) }is expressed as:
\begin{equation}
  \label{jredistrib}
  \tilde{J_\nu} = \frac{1}{\varphi_\nu} \int_0^\infty{R(\nu^\prime,\nu) J_{\nu^\prime} d\nu^\prime} .
\end{equation}
$\varphi_\nu$ is the normalized absorption profile of the line, and $R(\nu^\prime,\nu)$ is the angle-averaged frequency redistribution function. It gives the probability density that a photon absorbed at frequency $\nu^\prime$ is re-emitted at frequency $\nu$. 

{We follow the treatment of \cite{hummer62} who studied four types of redistribution of radiation scattered from moving atoms. The redistribution functions with the greatest significance \citep{mihalas} are Hummer's case~II (radiation damping with coherence in the atom's rest frame), and case~III (radiation and collision damping with complete redistribution in the atom's rest frame). A Doppler effect arises when deriving expressions for the redistribution functions in the observer's frame.
The case~II redistribution function $R_{II}$ is well suited to describe the scattering of radiation in resonance lines \citep{mihalas,cv78}, and we use it for the redistribution in the Lyman lines. In the Doppler core of the line (2--3 Doppler widths from line centre) this function is close to complete redistribution (CRD). Coherency effects are more noticeable in the line wings \citep{mm73}. This means that in the solar corona the coherency effects of $R_{II}$ may be important at about 1.5~\AA\ from the line centre of the Lyman lines, and thus difficult to detect.
The case~III redistribution function $R_{III}$ is close to CRD over the entire line profile. It is the predominant redistribution mechanism in collisionally-dominated regions of the solar atmosphere. 
For studies of the Lyman~$\alpha$ line profile in the corona, it is generally assumed that the redistribution function is well described in the coherent scattering approximation for an atom with two sharp energy levels (zero natural line width; \citealp[see e.g.,][]{witetal82apj,lietal98,loaetal98,cranmer98}). This is the case~I redistribution function in \citet{hummer62}. \cite{cranmer98} studied case~II and found that it makes a small difference relative to case~I which would be hardly noticed in the observations of the Lyman~$\alpha$ profile. This is due to the fact that both case~I and case~II redistribution functions are close to CRD in the Doppler core of the Lyman lines, which extend to about 1.5~\AA\ from line centre at coronal temperatures. The differences between case~I and case~II arise in the far wings of Lyman~$\alpha$ where the Thomson component of the line (chromospheric Ly$\alpha$ radiation scattered from electrons in the corona) becomes the major contributor to the line intensity. It would be interesting to see if this effect is similar for Lyman~$\beta$.}

{In this work we use partial redistribution in frequency (PRD) to compute the frequency redistribution function in Eq.~(\ref{jredistrib}) for all Lyman lines up to Ly-9, assuming isotropic scattering {in the laboratory frame}. Our redistribution function is therefore a linear combination of $R_{II}$ and $R_{III}$. The redistribution function $R_{III}$ is taken to be equal to the complete redistribution function given by the product $\varphi_\nu \varphi_{\nu^\prime}$ as in \cite{mihalas}.}
Defining the branching ratio $\gamma=\Gamma_r/(\Gamma_r+\Gamma_c)$, where $\Gamma_r$ and $\Gamma_c$ are the radiative and collisional damping constants respectively, we have {\citep{osc72}}: 
\begin{equation}
  R(\nu,\nu^\prime) = \gamma R_{IIA}(\nu,\nu^\prime) + (1-\gamma) \varphi_\nu \varphi_{\nu^\prime} \ .
  \label{defredi}
\end{equation}
We compute $\Gamma_c$ for each {Lyman} line at each position along the LOS. It is important to take this spatial variation into account as we found that it makes a difference in the width of the emergent line profiles if it is neglected. Furthermore, since the Lyman~$\alpha$ line has extended wings, we use a frequency-dependent collisional damping coefficient \citep{yeletal81}. All other subordinate lines are treated with the {standard} CRD approximation by imposing $\gamma=0$ in eq.~(\ref{defredi}).
A comparison between CRD and PRD computations shows that CRD alone would be a bad approximation for the Lyman lines. However CRD cannot be neglected, as we find {noticeable} differences in the line widths between {the scattering with $R_{II}$ only} {(by forcing $\gamma=1$)} and PRD for Lyman~$\beta$. 

\section{Results and discussion}\label{discuss}

\begin{figure}
  \centering
  \resizebox{\hsize}{!}{\includegraphics{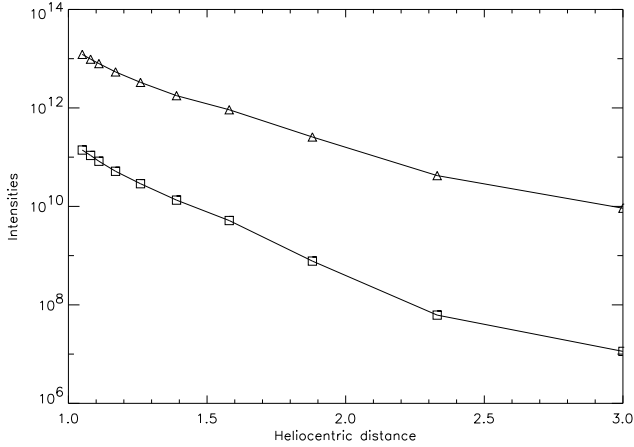}}
  \caption{Integrated intensities in photons s$^{-1}$ cm$^{-2}$ sr$^{-1}$ of Lyman~$\alpha$ (triangles) and Lyman~$\beta$ (squares) as a function of heliocentric distance.}
  \label{intensites}
\end{figure}

\begin{figure}
  \centering
  \resizebox{\hsize}{!}{\includegraphics{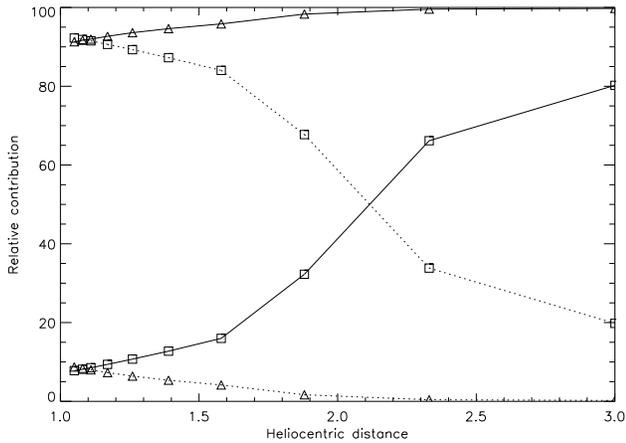}}
  \caption{Relative contribution of the radiative (solid line) and collisional (dotted line) components for Lyman~$\alpha$ (triangles) and Lyman~$\beta$ (squares) as a function of heliocentric distance.}
  \label{compo}
\end{figure}

Figure~\ref{intensites} presents the resulting integrated intensities of the first two Lyman lines as a function of distance from sun centre. The intensities are calculated by summing over the computed line profiles. The decrease of the intensity with altitude is more pronounced for Lyman~$\beta$ than for Lyman~$\alpha$, a first indication that they relate to the plasma parameters in different ways. The Lyman~$\alpha$ intensities compare well with the computed and observed values presented in \cite{vbr03}.
{We find that the variation of the Ly$\alpha$ intensity with height is not related to the variation of the electron density, while the decrease of Ly$\beta$ and H$\alpha$ (not shown on Fig.~\ref{intensites}) intensities follows the decrease of $n_e^2$ closely up to $\sim 1.5$~R$_{\sun}$, consistent with the fact that these two lines are mostly formed by collisional excitation in the inner corona. Then the fall-off in intensity is less rapid than that of the square of the electron density, owing to the growing importance of radiative excitation (see also Fig.~\ref{compo}).
In fact there is a coupling between Ly$\beta$ and H$\alpha$, which means that {an} H$\alpha$ photon can be absorbed and subsequently lead to an emission of a Ly$\beta$ photon. We obtain a nearly constant ratio between Ly$\beta$ and H$\alpha$ intensities, with $I(\mathrm{Ly}\beta)/I(\mathrm{H}\alpha) \simeq 8$. This coupling leads to lowered coherency effects, as was illustrated by \cite{hgv87}.
The coherence coefficient $\gamma$ (eq.~\ref{defredi}) is close to 1 for Ly$\alpha$ and 0.57 for Ly$\beta$ at the centre of the LOS closer to the Sun (height of 1.05~R$_{\sun}$).  We have computed the parameter $\lambda = (A_{ji} / P_j) \times \gamma$ as in \cite{hgv87} -- although we use a slightly different definition for $\gamma$, where $A_{ji}$ and $P_j$ are the spontaneous emission coefficient in the $j\to i$ transition and the total depopulation rate of level $j$, respectively. 
Close to the Sun, in the streamer base, our value of $\lambda$ is around 0.99 for Ly$\alpha$ and 0.3 for Ly$\beta$ (at the centre of LOS).  This confirms that the coherency effects in Ly$\beta$ are less important than in Ly$\alpha$.
}

Now we turn to the study of the Lyman line profiles. We fit each computed profile with a gaussian profile obtained from a non-linear least squares fit. 
{We find that the Ly$\alpha$ wings are broader than our gaussian fits at all altitudes. The line centre is well reproduced by the gaussian fits at altitudes above $\sim$1.6~R$_{\sun}$.
The Ly$\beta$ line deviates from the gaussian fits as the altitude increases. The line centre is always well reproduced by the gaussian fit, but the wings get broader with height.}
From our gaussian fits we obtain the full width at half maximum (FWHM) of the computed line profiles.  As we go further up in the corona, the line widths decrease. This reflects the decrease in temperature.
However, care should be taken when inferring the temperature from the line width. {As the profile is not exactly gaussian, this introduces an error in the temperature derivation \citep[for a more thorough discussion see, e.g.,][]{loaetal98}}. In this respect it seems more reliable to exploit the diagnostic possibilities of the Lyman~$\beta$ line which is closer to a gaussian {than the Lyman~$\alpha$ line}. 

\begin{figure}
  \centering
  \resizebox{\hsize}{!}{\includegraphics{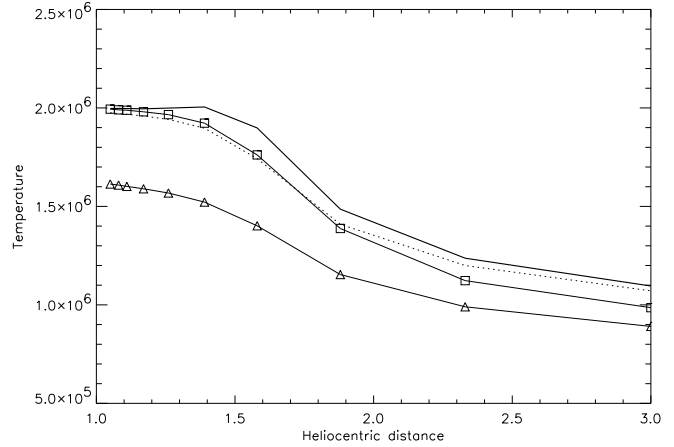}}
  \caption{Temperature derived from the width of Lyman~$\alpha$ (triangles), Lyman~$\beta$ (squares), temperature at centre of LOS (thick solid line), and  mean temperature from the model input (thick dotted line), as a function of heliocentric distance.}
  \label{temperatures}
\end{figure}

To infer the temperature of the neutrals from the line widths, one can relate the FWHM and the temperature $T_H$ with:
\begin{equation}
  T_H = \frac{m}{2k} \left[ \mathrm{FWHM}^2 \frac{c^2}{4 \lambda^2\ln 2}- \xi^2 \right] \ .
  \label{temp}
\end{equation}
We arbitrarily chose to use $\xi = 20$ km s$^{-1}$ for the non-thermal motions in the calculations of the line profiles at all altitudes. While it might not be accurate, the exact value for $\xi$ is not important for our discussion on the temperatures inferred from the Lyman lines, as we assume that they both have the same non-thermal broadening. The resulting temperatures derived from the width of the Lyman~$\alpha$ line and the  Lyman~$\beta$ line are plotted in Fig.~\ref{temperatures} as a function of height of the LOS, together with the mean temperature derived from the model input and the temperature at the centre of the LOS. The mean temperature is defined as $T_\mathrm{mean} = \int_0^M{T(m)\mathrm{d}m}/M$, with $m$ the column mass along the LOS, and $M$ the total column mass of the LOS.
Inspection of figure~\ref{temperatures} clearly shows that the width of the Lyman~$\beta$ line is indeed a good indicator of the plasma temperature from 1.05~R$_{\sun}$ up to $\sim$2~R$_{\sun}$. Above this height, the temperature derived from the Lyman~$\beta$ FWHM is slightly lower than the mean plasma temperature $T_\mathrm{mean}$ and than the central temperature, an indication that the plasma conditions have significantly changed. Indeed, from the streamer model the magnetic cusp is located at about 2~R$_{\sun}$. Within that distance, outflow velocities are very small (a few km~s$^{-1}$). Above the cusp, the protons reach a velocity of about 100 km~s$^{-1}$ at 3~R$_{\sun}$. 

Close to the Sun, the temperature derived from the Ly$\beta$ FWHM is in very good agreement with the temperature at the centre of the LOS. However, higher in the streamer, the temperature derived from the Ly$\beta$ line width is in better agreement with the mean plasma temperature than with the temperature at the centre of the LOS. 
Due to the temperature and hydrogen density variations along the LOS (see Fig.~\ref{streamer}), the contribution of collisional excitation in the formation of the Ly$\beta$ line is more concentrated at the centre of the LOS when close to the Sun, and more smoothly distributed along the line of sight higher in the streamer.
We note that when neglecting CRD in frequency redistribution in eq.~(\ref{defredi}), the temperature derived from the Ly$\beta$ width \emph{exactly} matches the mean temperature of the models up to a height of 1.6~R$_{\sun}$.
It can also be seen from Fig.~\ref{temperatures} that the temperature derived from the Lyman~$\alpha$ line significantly underestimates the mean plasma temperature and the temperature at the centre of the LOS. This difference can be as much as $3.8\times10^5$~K at $r=1.39$~R$_{\sun}$ for the mean temperature, and $5.0\times10^5$~K at $r=1.58$~R$_{\sun}$ for the central temperature. 
The difference in behaviour between the two Lyman lines points to the different relative contributions of radiative and collisional excitation in the formation of the two lines. From eq.~(\ref{source}) we identify the first term of the right-hand side with the collisional component and the second term with the radiative component. Figure~\ref{compo} shows their variation with altitude for the Lyman~$\alpha$ and Lyman~$\beta$ lines. The radiative component is represented with a solid line, and the collisional component with a dotted line, while triangles stand for Lyman~$\alpha$ and squares for Lyman~$\beta$.
This figure shows that, as expected, Lyman~$\beta$ is mostly formed by collisional excitation in the streamer. However it is interesting to note that the collisional component of Lyman~$\alpha$ is not negligible in the streamer base, contributing to nearly 10\% of the line intensity close to the Sun. We should stress here that both {the scattered and collisional} components will include contributions resulting from exchanges with other transitions (lines and continua) involving the rest of the atomic states.
The values reported in Fig.~\ref{compo} can be compared with those given by \citet{rayetal97} in the centre of a streamer observed by UVCS. These authors obtained a collisional contribution of 1.1\% for Lyman~$\alpha$ and 57\% for Lyman~$\beta$ at $\log T=6.2$, while our values at that temperature (at a heliocentric height of 1.7~R$_{\sun}$) lead to 3\% and 77\%, respectively. \\

In this paper NLTE radiative transfer calculations are performed to compute the properties of the Lyman lines of hydrogen in the solar corona. It is shown that the width of the Lyman~$\beta$ line is a better indicator of the plasma temperature than the width of the Lyman~$\alpha$ line, especially within the streamer. {It is due to the formation mechanisms of the lines, and the coupling of Ly$\beta$ with H$\alpha$. This work has been done using the approximation of isotropic frequency redistribution {in the laboratory frame}. We do not expect that the inclusion of angle-dependent redistribution functions in our calculations would change the main conclusion of this Letter, namely that the width of Lyman~$\beta$ is a better proxy for the plasma temperature in the streamer. It is known that the consideration of angular redistribution with dipole scattering has the effect of narrowing the line profile compared to isotropic redistribution, due to non-90$\degr$ scattering of photons \citep[e.g.][]{witetal82,loaetal98,cranmer98}. Therefore we believe that the discrepancy found here between temperatures derived from Ly$\alpha$ line profiles and model temperatures could be even greater by using angle-dependent redistribution functions. Furthermore, \cite{cv78} showed that the effects of angle-dependent PRD will be more important when the incident lines show substantial center-to-limb variations. This is not the case of hydrogen Lyman lines, but this is the case for, e.g., the \ion{O}{vi} lines at 1032 and 1038 \AA. We will include this effect in a future study when we include \ion{O}{vi} in our calculations.}
We also obtain new estimates of the radiative and collisional contributions of the Lyman line intensities in a non-isothermal streamer. These new values may have some importance in the derivation of element abundances. {Element abundances relative to hydrogen can be inferred independently from the ratio of the resonantly scattered (or collisional) component of a spectral line to the resonantly scattered (or collisional) component of, say, \ion{H}{i}~Ly$\beta$ \citep{witetal82,rayetal97}. Therefore the relative contributions of the two components of the Lyman lines to their total observed intensities must be known with good accuracy. Finally, }
our results can be compared with observations by the SUMER and UVCS spectrometers on SOHO. 
The radiative transfer calculations can be enhanced by including other effects such as Doppler dimming to improve the modelling in regions of the corona where outflow velocities cannot be ignored. This has been presented in \cite{corsoho17}.

\begin{acknowledgements}
  The authors are grateful {to the referee, Dr. P.~Heinzel, for his thorough comments that improved the clarity of this work, and} to P. Gouttebroze, S. Habbal and J.-C. Vial for their critical reading of an early version of the manuscript. Support from PPARC grant PPA/G/O/2003/00017 is acknowledged.
\end{acknowledgements}

\bibliographystyle{../../../LATEX/aa}
\bibliography{../../../LATEX/nicbib}

\end{document}